\documentclass[doublecol]{epl2} 
\usepackage{bm}
\usepackage{graphicx}
\usepackage{amssymb,amsfonts,amsmath}
\usepackage{subfigure}

\newcommand{\ri}{\mathrm{i}}

\newcommand{\rd}{\mathrm{d}}
\newcommand{\bo}{\hat{b}^{\phantom\dag}}
\newcommand{\ba}{\hat{b}^{\dag}}
\newcommand{\dop}{\hat{d}^{\phantom\dag}}
\newcommand{\da}{\hat{d}^{\dag}}
\newcommand{\no}{\hat{n}}
\newcommand{\Ho}{\hat{H}}
\newcommand{\Uo}{\hat{U}}
\newcommand{\xo}{\hat{\sigma}^{x}}
\newcommand{\yo}{\hat{\sigma}^{y}}
\newcommand{\zo}{\hat{\sigma}^{z}}
\newcommand{\phio}{\hat{\varphi}}
\newcommand{\la}{\langle}
\newcommand{\ra}{\rangle}
\newcommand{\be}{\begin{equation}}
\newcommand{\ee}{\end{equation}}
\newcommand{\bes}{\begin{eqnarray}}
\newcommand{\ees}{\end{eqnarray}}

\newcommand{\br}{{\bm r}}
\newcommand{\bF}{{\bm F}}

\newcommand{\bq}{{\bm q}}
\newcommand{\bz}{{\bm 0}}

\newcommand{\bp}{{\bm p}}
\newcommand{\bx}{{\bm x}}
\newcommand{\bfe}{{\bm e}}
\newcommand{\bfa}{{\bm a}}
\newcommand{\rJ}{\mathrm{J}}
\title{Frustrated quantum antiferromagnetism with ultracold bosons in a
triangular lattice}
\shorttitle{Frustrated quantum antiferromagnetism with ultracold bosons}
\author{
A.\ Eckardt\inst{1}\thanks{E-mail:\email{andre.eckardt@icfo.es}}
\and P.\ Hauke\inst{1}
\and P.\ Soltan-Panahi\inst{2}
\and C.\ Becker\inst{2}
\and K.\ Sengstock\inst{2}
\and M.\ Lewenstein\inst{1,3}}

\shortauthor{Eckardt et al.}

\institute{                    
  \inst{1} ICFO-Institut de Ci\`encies Fot\`oniques, 
	Av.\ Canal Ol\'impic s/n, E-08860 Castelldefels (Barcelona), Spain   
  \\
  \inst{2} Universit\"at Hamburg, Luruper Chaussee 149, 
					D-22761 Hamburg, Germany
  \\
  \inst{3} ICREA-Institucio Catalana de Recerca i Estudis Avan\c{c}ats,
Lluis Companys 23, E-08010 Barcelona, Spain
}


\abstract{We propose to realize the anisotropic triangular-lattice Bose-Hubbard model with
positive tunneling matrix elements by using ultracold atoms in an optical
lattice dressed by a fast lattice oscillation. This model exhibits frustrated
antiferromagnetism at experimentally feasible temperatures; it interpolates
between a classical rotor model for weak interaction, and a quantum spin-1/2
$XY$-model in the limit of hard-core bosons. This allows to
explore experimentally gapped spin liquid phases predicted recently
[Schmied et al., New J.\ Phys.\ {\bf 10}, 045017 (2008)].}

\begin{document}

\maketitle

\section{Introduction}

Frustrated quantum antiferromagnetism can give rise to
extraordinarily rich physics \cite{Reviews}. It is not only supposed to play
a crucial role for the properties of high-$T_c$ superconductors, but it is also
an interesting subject on its own, and has
potential applications in topological quantum information processing and
storage \cite{NayakEtAl08}. Apart from the possibility of keeping a classical
N\'eel-ordered (staggered or spiral) spin configuration with long-range 
order, the spins of a quantum antiferromagnet can also form singlet pairs
(``valence bonds''), such that the spin-rotation symmetry is not broken.
These singlets either order spatially to form a valence bond solid
or the system's state is a superposition of many singlet coverings, neither
breaking translational nor spin-rotational symmetry. The latter is termed
a resonating valence bond spin liquid (SL).
A gapped SL with an exponential decay of spin correlations is expected to
exhibit non-local topological order being immune against local perturbations.
It also can possess anyonic excitations. This makes such \emph{topological} SL
states candidates for robust quantum memories and processors \cite{NayakEtAl08}.
Alternatively, a \emph{critical} SL is characterized by a huge density of
low-lying excitations and a power-law decay of spin correlations.

Since frustrated quantum antiferromagnets are hard to simulate (path-integral
Monte-Carlo methods fail) and clean solid-state realizations are not
available, it is desirable to study these exotic many-body systems with
ultracold atoms in optical lattice potentials \cite{OpticalLattices} providing
both clean conditions and far-reaching control. 
The most straightforward cold atom implementation of a quantum magnet is to
create a Mott-insulator of fermions in two different internal states forming a
pseudo spin. However, the necessary low temperatures (smaller than the weak
superexchange spin coupling) have not yet been achieved. Also spinless fermions
at filling 2/3 in a not yet realized trimerized Kagom\'e lattice resemble a
quantum magnet \cite{DamskiEtAl05}. 

In this letter we propose a different strategy for the realization of a
frustrated quantum system with ultracold atoms that --- in contrast to the
aforementioned approaches --- can be pursued in existing experimental setups,
at temperatures already reached. Our idea is to consider spinless ultracold
bosonic atoms in a triangular optical lattice and to induce frustration via a
sign change of the matrix elements describing tunneling between adjacent
potential minima. As will be shown, such a sign change can be achieved
effectively by dressing the system with a high-frequency elliptical lattice
acceleration.
In the hard-core boson limit of strong repulsive interaction, the physics is
then described by the antiferromagnetic spin-1/2 $XY$-model on the triangular
lattice. For certain regimes of anisotropic coupling this model is expected to
show gapped SL phases \cite{SchmiedEtAl08}. 

This letter is organized as follows. We start with a discussion of the
frustrated positive-hopping Bose-Hubbard model that describes the system to be
realized experimentally. In order to sketch the
expected phase diagram we combine (i) recently published
numerical data (based on PEPS as well as exact simulations) \cite{SchmiedEtAl08}
valid in the limit of strong interaction with (ii) results obtained by starting
from the limit of weak interaction and systematically including quantum
fluctuations (beyond Bogoliubov).
Then we discuss the experimental realization of the model, putting
emphasis on how to change the sign of the tunneling matrix elements via a fast
elliptical lattice acceleration. It follows a part devoted to the preparation of
the frustrated model's ground state in the presence of a trapping potential.
Finally, before giving a brief conclusion, we discuss possible experimental
signatures of the expected phases.

\section{Positive-hopping Bose-Hubbard model on a triangular lattice}
\begin{figure}[t]\centering
\includegraphics[width = 1\linewidth]{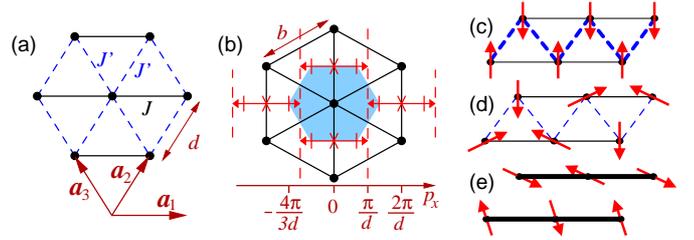}
\caption{\label{fig:lattice} (color online) (a) Anisotropic triangular lattice
considered, with primitive vectors
$\bfa_1\equiv d\bfe_x$, $\bfa_2\equiv d[(1/2)\bfe_x+(\sqrt{3}/2)\bfe_y]$, as
well as $\bfa_3\equiv -\bfa_1+\bfa_2$. The tunneling matrix elements 
$J_{ij}^\text{eff}$ take values $J$ and $J'\equiv\alpha J$ (with $\alpha\ge0$) 
along the solid and dashed bonds, respectively.
(b) Reciprocal lattice with $b=(4\pi/\sqrt{3})d^{-1}$. The
first Brillouin-zone, centered at $\bp=\bz$, is shaded.
Considering antiferromagnetic coupling $J>0$, we have marked the ordering vector
$\bq$ describing a N\'eel SF in the limit of weak interaction: For 
$\alpha\ge\alpha_0$ $\bq$ lies on one of the x-shaped crosses (being 
equivalent modulo reciprocal lattice vectors). This corresponds to a staggered 
configuration of the local phase angles $\varphi_i$ on the rhombic lattice of 
$J'$-bonds [shown in (c) with the $\varphi_i$ visualized by pointers]. Lowering
$\alpha$, at $\alpha=\alpha_0$ the position of $\bq$ splits in a continuous way
into two non-equivalent possible positions that 
separate symmetrically along the arrows drawn in (b). The phases $\varphi_i$
assume a spiral pattern with two possible chiralities;
subfigure (d) corresponds to the isotropic lattice ($\alpha=1<\alpha_0$) with $\bq$ lying
on one of the corners of the first Brillouin zone. 
Finally, in the 1D limit
($\alpha=0$) only $q_x$ has a defined value that is marked by the dashed lines in
(b). The phase pattern is staggered along the 1D chains of $J$-bonds [as sketched in (e)].}
\end{figure}

Consider a sample of ultracold bosonic atoms in a deep triangular 
optical lattice that is forced inertially by moving the lattice rapidly 
along an elliptical orbit. According to the following section, the 
time evolution of such a system has a simple description.
Integrating out the fast oscillatory motion on the short time scale
$T=2\pi/\omega$ of the elliptical forcing, one finds the
system's evolution on longer time scales governed by the time-independent
effective Bose-Hubbard Hamiltonian
	\be\label{eq:Heff}
	\Ho_\text{eff} = \sum_{\la ij\ra}J^\text{eff}_{ij}
	\ba_i\bo_j
		+\sum_i \left[\frac{U}{2}\no_i(\no_i-1)-\mu_i\no_i\right].
	\ee
Here $\hat{b}_i$ and $\no_i$ are the bosonic annihilation and number
operators for Wannier states localized at the minima $\br_i$ of the
triangular lattice potential. The first term comprises tunneling between
adjacent sites $i$ and $j$ with --- this is the crucial point --- matrix
elements $J_{ij}^\text{eff}$ that are smoothly tunable from negative to positive
values by variation of the forcing strength.\footnote{In our convention
$\la ij\ra$ denotes an oriented pair of neighboring sites,
$\la ij\ra\ne\la ji\ra$.}
The on-site terms are characterized by the positive interaction parameter $U$
and the local chemical potential $\mu_i\equiv\mu-V_i$ including the trapping
potential $V_i$. We consider the anisotropic lattice shown in
Fig.~\ref{fig:lattice}(a) with the $J_{ij}^\text{eff}$ equal to either $J$ or
$J'\equiv\alpha J$ (assuming $\alpha\ge0$). 

\begin{figure}[t]\centering
\includegraphics[width = 1\linewidth]{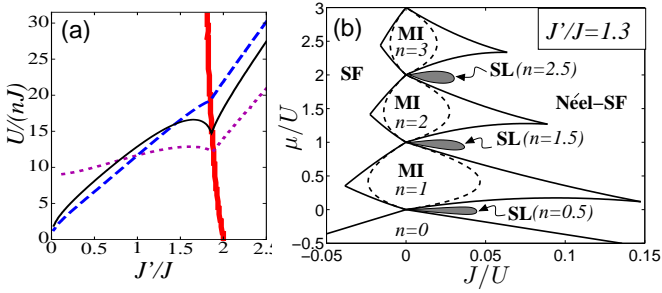}
\caption{\label{fig:app} (color online) 
(a) Generalized Bogoliubov theory for $n=3.5$. Values of
$U/(n|J|)$ at which: spiral changes to rhombic-staggered N\'eel order (thick line),
relative phase fluctuations %$\la(\delta\phio_i-\delta\phio_j)^2\ra^{\!1/2}$
reach $\pi/4$ for sites separated by $\bfa_1$ (dotted line) and $\bfa_2$
(dashed line), the condensate fraction %$f_\text{c}$ 
has dropped to $0.75$ (thin solid line).
(b) Boundaries of the MI phases with
integer filling $n$ in the $\mu/U$-$J/U$-plane, both in 2nd-order 
strong-coupling (solid lines) and meanfield (dashed lines) approximation. As a
consequence of frustration, the MI double-lobes are larger on the antiferromagnetic
side ($J>0$) of the phase diagram. The grey bubbles between the
MI regions, indicating the expected gapped SL phases at half-odd-integer
filling, are just sketched. 
%For large negative (positive) $J/U$ the system is a ordinary SF (N\'eel SF).
} 
\end{figure}

The homogeneous model ($\mu_i=\mu$) interpolates between a classical rotor and a
quantum spin model: For weak interaction $U\ll n|J|$, with a mean filling of $n$ 
particles per site, the superfluid (SF) ground state can (locally) be approximated by
$\prod_i\exp(\psi_i\ba_i)|\text{vacuum}\ra$ with discrete order parameter 
$\psi_i =\sqrt{n_i}\exp(\ri\varphi_i)$. A homogeneous density $n_i=n$ is
favored and the local phases $\varphi_i$ play the role of classical rotors
assuming a configuration $\varphi_i\equiv\bq\cdot\br_i$ described by the ordering
vector $\bq$. {Antiferromagnetic coupling $J>0$ implies N\'eel ordered 
phases $\varphi_i$ as depicted in Fig.~\ref{fig:lattice}(b-e). We call such a state a
 N\'eel SF. When $\alpha$ exceeds a value $\alpha_0$, spiral continuously transforms
into staggered N\'eel order. While $\alpha_0$ equals 2 for $U/J=0$, it 
slightly decreases with increasing interaction, cf.\ Fig.~\ref{fig:app}(a).}

\begin{figure}[t]\centering
\includegraphics[width = 0.9\linewidth]{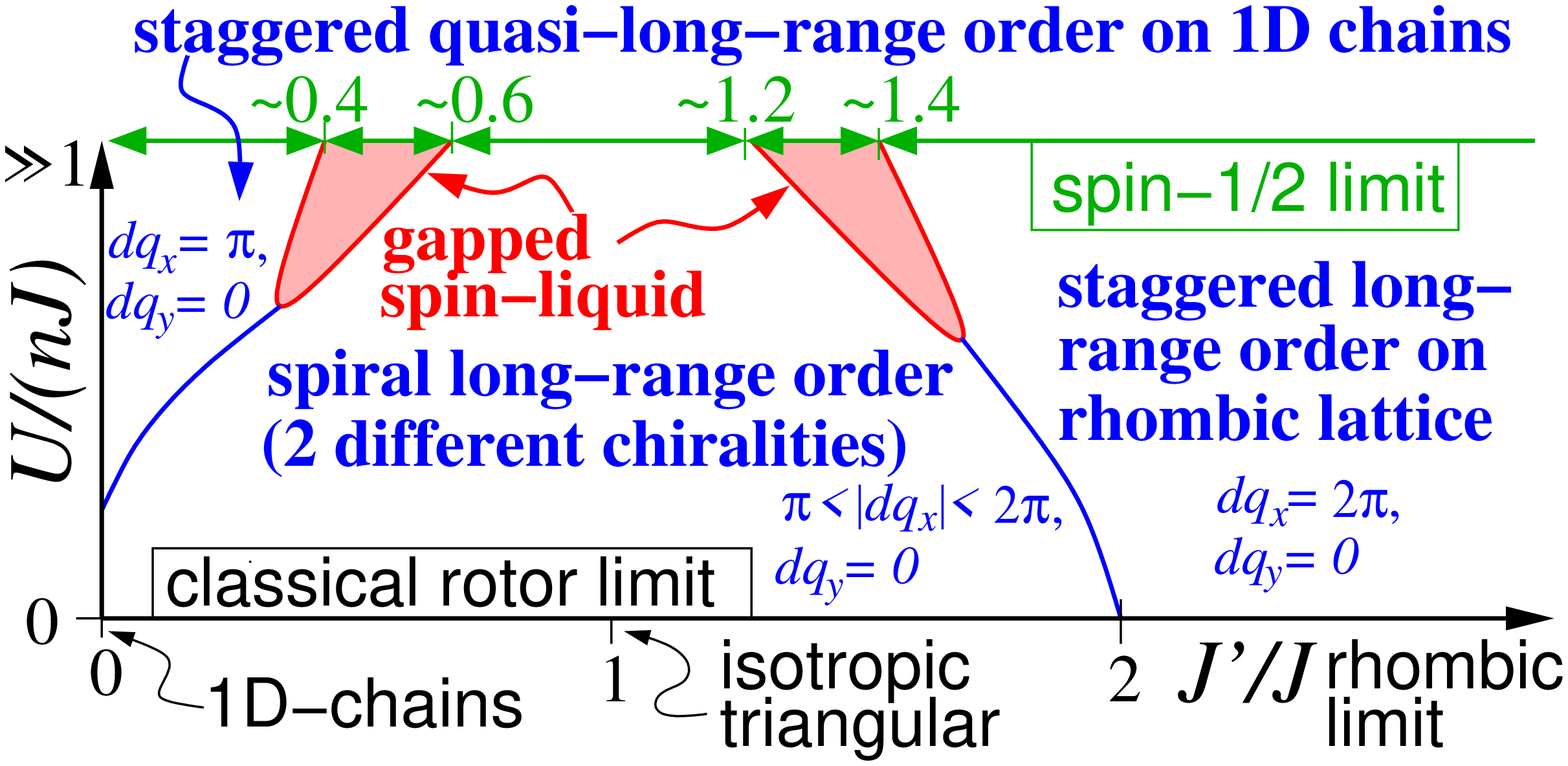}
\caption{\label{fig:diagram}(color online) Sketch of the phase diagram of the
anisotropic positive-hopping Bose-Hubbard model on the triangular lattice for
half-odd-integer filling. The parameter plane is spanned by interaction strength
$U/(nJ)$ and anisotropy ratio $\alpha=J'/J$. The data (in green) for the
spin-1/2 limit [$U/(nJ)\gg1$] are taken from Ref.~\cite{SchmiedEtAl08}. We
assume the SL phases to survive at small finite values of $J/U$, since they are
protected by a gap. The behaviour at small $U/(nJ)$ corresponds to
results obtained within a generalized Bogoliubov theory,
cf.~Fig.~\ref{fig:app}(a).}
\end{figure}

In the opposite limit of strong interaction $U\gg n|J|$, there are only two
energetically favored site occupations, $n_i=[n]\equiv g$ (the largest
integer smaller than $n$) and $n_i=g+1$. Associating them with ``spin up'' and
``spin down'', respectively, gives the Bloch-sphere representation 
$|\vartheta_i,\varphi_i\ra\equiv\cos(\vartheta_i/2)|g\ra_i 
+\sin(\vartheta_i/2)\exp(\ri\varphi_i)|g+1\ra_i$ at each site. Replacing
$(g+1)^{\!-1}\bo_i$ by the spin lowering operator \mbox{$(\xo_i-\ri\yo_i)/2$}, one
arrives at the $XY$-model 
	\be\label{eq:Hxy}
	\Ho_{XY} = 
		\sum_{\la i j\ra}J^{XY}_{ij} (\xo_i\xo_j+\yo_i\yo_j)
		+\sum_i h_i \zo_i
	\ee
with $h_i\equiv\frac{1}{2}(\mu_i-Ug)$, $J^{XY}_{ij}\equiv
\frac{g+1}{4}J^\text{eff}_{ij}$, and $\xo_i$, $\yo_i$, $\zo_i$ being spin-1/2
Pauli operators at site $i$. The ground state of $\Ho_{XY}$ cannot be a product
state like $\prod_i|\vartheta_i,\varphi_i\ra$ with definite local phases
$\varphi_i$ anymore, since $|\vartheta_i,\varphi_i\ra$ cannot be an eigenstate
of both $\xo_i$ and $\yo_i$. Viewed from the Bose-Hubbard perspective, increasing
interparticle repulsion increases the fluctuations of the local phases $\varphi_i$.
While for ferromagnetic coupling $J<0$ the classical phase configuration is
supposed to survive the presence of quantum fluctuations in the spin-1/2 limit
$U\gg n|J|$, for antiferromagnetic coupling $J>0$ recent simulations suggest
that (for $\sum_i\la\zo_i\ra=0$) classical N\'eel order is not necessarily
preserved \cite{SchmiedEtAl08}: Along the $\alpha$-axis different N\'eel phases
are separated by gapped SL phases with exponentially decaying spin correlations.
The results of Ref.~\cite{SchmiedEtAl08} are displayed along the upper edge of
the phase diagram shown in Fig.~\ref{fig:diagram}.

In order to gain further insight into the physics of the frustrated
positive-hopping Bose-Hubbard model (\ref{eq:Heff}), we start from the
classical limit of weak interaction $U\ll n|J|$ (assuming a homogeneous system) and
include quantum fluctuations by using the generalized Bogoliubov approach
introduced in Ref.~\cite{MoraCastin03}. For filling already moderately larger
than 1, we can replace
$\bo_i\simeq \exp[\ri(\varphi_i+\delta\phio_i)]\sqrt{n_i+\delta\no_i}$, where
$\delta\no_i=\delta\no_i^\dag$ and $\delta\phio_i\simeq\delta\phio_i^\dag$
describe quantum fluctuations to the local particle numbers $n_i$ and 
phases $\varphi_i$, respectively, with
$[\delta\no_i,\delta\phio_i]\simeq\ri\delta_{i,j}$. While
$\la(\delta\no_i)^2\ra_i/n^2\ll1$ can be assumed, the phase fluctuations
$\la(\delta\phio_i)^2\ra$ diverge in the 1D limit ($\alpha=0$) (as well as at
finite temperatures) where only quasi-long-range order is possible. However,
one can still expect the fluctuation of the \emph{relative} phases
$\la(\delta\phio_i-\delta\phio_j)^2\ra$ between neighboring sites $i$ and $j$ to
be small. Expanding the Hamiltonian (\ref{eq:Heff}) up to second order in
$\delta\no_i/n_i$ and $(\delta\phio_i-\delta\phio_j)$, it will be quadratic in
terms of new bosonic operators
$\dop_i\equiv\sqrt{n_i}[\delta\no_i/(2n_i) + \ri\delta\phio_i]$ and $\da_i$, and
can be diagonalized by a Bogoliubov transform (keeping $\la\delta\no_i\ra=0$).
When computing, e.g., correlations $\la\ba_i\bo_j\ra$ between distant sites $i$
and $j$, one cannot treat $(\delta\phio_i-\delta\phio_j)$ as a small quantity,
but has to use Wick's theorem to evaluate expectation values of all powers of
$\delta\phio_i$ \cite{MoraCastin03}. We augment this analysis by taking into account 
also the (Wick-decomposed) quartic corrections to the Hamiltonian when minimizing 
the ground-state energy with respect to both the Bogoliubov coefficients and the
ordering vector $\bq$.\footnote{In the case of spiral order ($\pi<d|q_x|<2\pi$
with $q_y=0$), one finds two solutions, $\bq$ and $\bq'=-\bq$, and has to choose
one of them.} This self-consistent above-Bogoliubov correction, that we include
using a numeric  iteration scheme, is necessary in order to explain a shift of
$\alpha_0$ with increasing interaction.\footnote{
Taking into account self-consistently the quartic terms does not lead 
to a spurious gap in the quasiparticle spectrum, as it is the case within the 
standard Bogoliubov treatment \cite{Griffin96}. This gaplessness 
is, thus, a feature of the generalized Bogoliubov expansion \cite{MoraCastin03} in 
terms of fluctuations in density and relative-phase.}

Assuming homogeneous filling $n_i=n$ as well as $q_y=0$, the method sketched
in the above paragraph leads to the following results:
With increasing interaction/quantum fluctuations, $\alpha_0$
decreases, i.e.\ the $\alpha$-domain of  rhombic-staggered N\'eel order grows
[thick line in Fig.~\ref{fig:app}(a)] . This {``order by disorder''
phenomenon  \cite{Reviews}} is in accordance with the spin-1/2 results of
 Ref.~\cite{SchmiedEtAl08} (cf.\ upper edge of Fig.~\ref{fig:diagram}). In contrast,
a finite $\alpha$-interval of staggered 1D quasi-long-range N\'eel order, also predicted for the spin
model, is not found. We have used these findings to draw the lower part of the
phase diagram of Fig.~\ref{fig:diagram}.
The quasiparticle dispersion relation is gapless and phonon-like
for quasimomentum wave numbers $\bp$ around $\bp=\bq$.
%\footnote{{
%At $\alpha=\alpha_0$, where spiral continuously transforms into staggered N\'eel
%order, the dispersion relation softens in $p_x$-direction. Without
%quartic corrections it becomes quadratic in $|p_x-q_x|$; including these
%corrections, however, leads to a small contribution linear
%in $|p_x-q_x|$ remaining at $\alpha=\alpha_0$.}}
However, whenever spiral order is found, it is symmetric with respect to 
$(\bp-\bq)\to-(\bp-\bq)$ only in the limit of small $|\bp-\bq|$.
In contrast, the zero temperature quasimomentum distribution, having sharp peaks at
$\bp=\bq$ + reciprocal lattice vectors, possesses reflection symmetry with respect
to $\bp=\bq$. 
Further results, for $n=3.5$, are shown in  Fig.~\ref{fig:app}(a). 
An estimate for the range of validity of the approximation is given by the 
dotted and the dashed line. Above them the relative phase fluctuations 
between neighboring sites separated by $\bfa_1$ and $\bfa_2$, respectively,
exceed a value taken to be $\pi/4$. The fact that the dotted line does not
approach zero in the limit of decoupled 1D chains ($\alpha\to0$) indicates that
in this limit the approximation still captures the physics in
$\bfa_1$-direction (along the chains). Moreover,
the dip around $\alpha=\alpha_0$ can be interpreted as a precursor of the SL
phase predicted in Ref.~\cite{SchmiedEtAl08} (cf.~Fig.~\ref{fig:diagram}).
Finally, the thin solid line marks the interaction where the condensate fraction
$f_c\equiv\lim_{|\br_{ij}|\to\infty} |\la\ba_i\bo_j\ra|/n$ is reduced 
to 0.75. Again, a sharp dip at $\alpha=\alpha_0$ is a hint at a quantum
disordered phase in the limit of large interaction.

\section{{Proposal for an} experimental realization}

Having discussed the physics of the triangular-lattice positive-hopping
Bose-Hubbard Hamiltonian, let us turn to the realization of the
model with ultracold atoms in a deep optical lattice. The sign change of the
tunneling matrix element, from negative to positive values, shall be induced by
dressing the system with a fast time-periodic lattice acceleration. 
For hypercubic lattices, such a dynamical modification of tunneling
has been predicted theoretically not only for single \cite{DynamicLocalization},
but also for many interacting particles \cite{EckardtEtAl05II}. Moreover, it has
been observed experimentally with ultracold atoms both in the weakly interacting
regime (via the expansion of a Bose-Einstein condensate \cite{LignierEtAl07}),
as well as in the strong-coupling regime, where it has been used to induce the
quantum phase transition from a SF to a Mott insulator (MI) and back
\cite{EckardtEtAl05II,ZenesiniEtAl09}. However, the linear driving scheme
used in the work just mentioned, with the system being forced sinusoidally along
a single direction (chosen to be diagonal with respect to all symmetry axes in
the case of a square or a cubic lattice), is not suitable for the triangular
lattice geometry. In order to be able to manipulate the system in a symmetric
way with respect to the three non-orthogonal lattice directions $\bfa_1$,
$\bfa_2$, and $\bfa_3$ [Fig.~\ref{fig:lattice}(a)], here we propose to use
elliptical forcing. This includes isotropic circular as well as linear forcing. 

The driving scheme to be considered can be realized inertially by
moving the lattice along an elliptical orbit
$\bx(t)=\Delta x_c\cos(\omega t)\bfe_c+\Delta x_s \sin(\omega t)\bfe_s$ in
space, with angular frequency $\omega$, orthogonal unit vectors $\bfe_c$ and 
$\bfe_s$, as well as amplitudes $\Delta x_c$ and $\Delta x_s$.
The resulting inertial force in the lattice frame of reference reads 
$\bF(t)= -m\ddot{\bx} = F_c\cos(\omega t)\bfe_c + F_s\sin(\omega t)\bfe_s$
where $m$ is the boson mass and $F_{c/s}=m\omega^2\Delta x_{c/s}$. Choosing
$\omega$ and $F_{c/s}$ small enough to exclude transitions from the lowest to
higher Bloch bands, one can describe the system in the lattice frame of
reference by the driven Bose-Hubbard model
	\be\label{eq:dbh}
	\Ho(t) = \sum_{\la ij\ra}J_{ij}\ba_i\bo_j  
		 +\frac{U}{2}\sum_i \no_i(\no_i-1)
		 + \sum_i[v_i(t)-\mu_i]\no_i.
	\ee
Here $J_{ij}<0$ are the bare tunneling matrix elements and
$v_i(t)\equiv-\br_i\cdot\bF(t)$ oscillating on-site energies.

We assume that $\hbar\omega$ is large compared to the energy scales given 
by interaction ($U$), tunneling ($n|J_{ij}|$), and trapping ($|\mu_i-\mu_j|$, with 
neighbors $i$ and $j$), i.e.\ that the forcing is fast with respect to the time
scales governing the undriven model. Under these conditions the time evolution of 
the driven system's state $|\psi(t)\ra$ will be to good approximation of the form 
	\be
	|\psi(t)\ra \approx \Uo(t)|\psi_\text{eff}(t)\ra.
	\ee
The unitary operator 
	\be
	\Uo(t)\equiv \exp\Big(-\frac{\ri}{\hbar}\sum_i \no_i W_i(t)\Big),
	\ee
where
	\be
	W_i(t)\equiv\int_0^t\!\rd\tau\,v_i(\tau)
-\frac{1}{T}\int_0^T\!\rd t'\int_0^{t'}\!\rd\tau\,v_i(\tau),
	\ee	
just describes a periodically time-dependent shift by $-m\dot{\bx}$ of the
whole system in quasimomentum, $W_i=\br_i\cdot m{\dot{\bx}}$. On top of this
simple oscillatory motion on the short time scale $T=2\pi/\omega$, the
time evolution on longer times is governed by the effective time-independent
Hamiltonian $\Ho_\text{eff}$ shown in Eq.~(\ref{eq:Heff}), namely
	\be
	|\psi_\text{eff}(t)\ra
		=\exp\bigg(-\frac{\ri}{\hbar}\Ho_\text{eff}t\bigg)
			|\psi_\text{eff}(0)\ra.
	\ee
The dressed tunneling matrix elements are given by
	\be\label{eq:Jeff}
	J_{ij}^\text{eff}=
			 J_{ij}\rJ_0\bigg(\frac{K_{ij}}{\hbar\omega}\bigg).
	\ee 
Here $\rJ_0$ is the zero-order Bessel function and
$K_{ij}\equiv\sqrt{(F_c\bfe_c\cdot\br_{ij})^2+(F_s\bfe_s\cdot\br_{ij})^2}$ 
the amplitude of the potential modulation between site $i$ and $j$, where  
$\br_{ij}\equiv\br_i-\br_j$. Thus, in the lattice frame, apart 
from the superimposed fast oscillation in quasimomentum, the system behaves as 
the one described by $\Ho_\text{eff}$.
When measuring the momentum distribution 
of the system in the laboratory frame by taking time-of-flight absorption
images, one will encounter the periodic quasimomentum distribution of
$|\psi_\text{eff}\ra$ at rest, being enveloped by the momentum distribution
of the Wannier wave function oscillating like $m\dot{\bx}$.

The result presented in the preceding paragraph relies on the separation of
time scales as well as on time averaging. We have obtained it within the
framework of quantum Floquet theory \cite{Floquet} by generalizing the approach
introduced in Refs.~\cite{EckardtEtAl05II,EckardtHolthaus08b} to elliptical
forcing. The derivation is based on stationary degenerate-state perturbation
theory on the level of an extended Hilbert space including time as a coordinate.
Here
we just give a simple argument making the $\Ho_\text{eff}$-description
plausible: Transforming $|\psi'\ra=\Uo^\dag|\psi\ra$ leads to the
new Hamiltonian $\Ho'=\Uo^\dag\Ho\Uo-\ri\hbar\Uo^\dag(\rd_t\Uo)$. Accordingly,
$\Ho'$ is obtained from $\Ho$ by subracting the oscillating potential terms 
$\propto v_i(t)$ and replacing $J_{ij}\to J_{ij}\exp(\ri[W_i-W_j]/\hbar)$.
Now the rapidly oscillating phase factors in the tunneling terms of $\Ho'$ can
approximately be taken into account on time average,
$\Ho'(t)\to\frac{1}{T}\int_0^T\!\rd t\,\Ho'(t)=\Ho_\text{eff}$, giving
$|\psi'\ra\approx|\psi_\text{eff}\ra$.

A 2D triangular optical lattice can be realized by superimposing three laser 
beams, all polarized in $z$-direction, at an angle of $2\pi/3$ in the
$xy$-plane, while a standing light wave in $z$-direction is used to create a
stack of effectively two-dimensional systems. A further beam in $z$-direction
allows to modify the trapping potential in the $xy$-plane. The lattice motion
can be realized by varying the relative frequencies of the beams by means of 
acousto-optical modulators. For the purposes described above, an orbit
$\bx(t)=\Delta x_c\cos(\omega t)\bfe_c+\Delta x_s \sin(\omega t)\bfe_s$ is
required, with $\Delta x_{s/c}$ on the order of a lattice constant and
$\omega/(2\pi)$ being a few kHz.  
Starting from an isotropic undriven lattice with bare tunneling matrix
elements $J_{ij}=\bar{J}<0$ and choosing $\bfe_{c/s}=\bfe_{x/y}$,
one obtains effective tunneling matrix elements (\ref{eq:Jeff}) distributed as
depicted in Fig.~\ref{fig:lattice}(a). Namely $K_{ij}$ reads
$K\equiv d|F_c|$ and $K'\equiv d\sqrt{F_c^2+3F_s^2}/2$ along the solid and
dashed bonds, respectively, giving $J=\bar{J}\rJ_0\big(K/(\hbar\omega)\big)$ and
$J'=\bar{J}\rJ_0\big(K'/(\hbar\omega)\big)$ according to Eq.~(\ref{eq:Jeff}).
This allows for any value of the anisotropy parameter $\alpha=J'/J$.

{We have already implemented a triangular optical lattice in
the laboratory, loaded it with ultracold $^{87}$Rb atoms, and observed the transition 
from a SF to a MI. Also a controlled motion of the lattice 
has been achieved.}

\section{State preparation and role of trapping potential}
For elliptical forcing there are no instants in time where $\Uo(t)$ is equal 
to the identity (i.e.\ with $\dot{\bx}=0$) like it is the case for linear
forcing ($F_s=0$) at integer $t\omega/(2\pi)$.
Thus, it is not possible to ``map'' the state $|\psi\ra$
of an initially unforced system on $|\psi_\text{eff}\ra$ by suddenly switching
on the forcing. However, one can smoothly switch on the drive. According to
the adiabatic principle for quantum Floquet states \cite{BreuerHolthaus89II},
$|\psi_\text{eff}\ra$ can follow adiabatically when $\Ho_\text{eff}$ is
modified by the forcing, starting from $|\psi_\text{eff}\ra=|\psi\ra$ in the
undriven limit \cite{EckardtEtAl05II,EckardtHolthaus08b}. 
Before passing from the ground state of the undriven system 
($J_{ij}^\text{eff}=J_{ij}<0$) to the positive-hopping regime 
($J_{ij}^\text{eff}>0$) in the presence of a trapping potential, the lattice
should be tuned very deep, such that $U\gg n_0|J|$ with filling $n_0$ in the
trap center. The system will form MI regions \cite{Mott}
with an integer number $g$ of particles (depending on $\mu_i/U$) localized at
each site. Different MI regions will be separated only by tiny intermediate
domains of non-integer filling. In the MI phases the state can follow smoothly
through the sign-change of $J$ when the lattice acceleration is ramped up
continuously in a next step. Moreover, in a deep lattice unwanted interband
transitions are strongly suppressed. When the desired strength of the 
forcing is reached, in the center of the trap the MI has to be melted. This can be 
achieved both by decreasing the lattice depth (without leaving the regime of strong
correlation $U\sim n_0|J|$) and by tuning the chemical potential in the trap center. 
The latter can be achieved by varying the trap, such that atoms are pushed into or 
pulled out of the center.

We have studied the MI-to-SF ($J<0$) and MI-to-N\'eel SF ($J>0$) transition in
the triangular lattice theoretically.
In the parameter plane spanned by $\mu/U$ and $J/U$, a strong-coupling expansion as 
described in Ref.~\cite{FreericksMonien94} gives the upper and
lower boundary, $\mu_\text{p}/U$ and $\mu_\text{h}/U$, of the MI phase with
integer filling $n=g$. One finds
$\mu_\text{p}/U=g -(g+1)\eta-gc_\text{p}\eta^2+\mathcal{O}(\eta^3)$ and
$\mu_\text{h}/U= (g-1)+g\eta+(g+1)c_\text{h}\eta^2+\mathcal{O}(\eta^3)$. The
expansion parameter is given by $\eta\equiv-\varepsilon(\bq)/U= w|J|/U$ with 
$\varepsilon(\bq)\equiv-|J|w$ being the single-particle dispersion relation 
$\varepsilon(\bp)\equiv 2J[\cos(dp_x)+2\alpha\cos(dp_x/2)\cos(\sqrt{3}dp_y/2)]$
evaluated at its minimum $\bq$. The coefficients read $c_\text{p} \equiv g+1 - 
(5g+4)(1+2\alpha^2)/w^2$ and $c_\text{h} \equiv g - (5g+1)(1+2\alpha^2)/w^2$. 
Here $w$ directly reflects frustration; while $w=4\alpha+2$ for
ferromagnetic $J<0$, it is smaller for antiferromagnetic $J>0$, namely 
$w=\alpha^2+2$ for $0\le\alpha\le2$ and $w=4\alpha-2$ for $\alpha\ge2$.
As a consequence, the MI regions extend to larger values of $|J|/U$ on the 
frustrated side of the $J-\mu$ plane. This can also be observed in 
Fig.~\ref{fig:app}(b) displaying the phase diagram for $\alpha=1.3$. Moreover, the 
transition from 1D like concave phase boundaries $c_\text{p/h}<0$ to square lattice 
like convex ones $c_\text{p/h}>0$ happens at noticeably larger $\alpha$ in the case
of fustration. Namely it occurs for 
$\alpha$ between 0.03 and 0.13  (2.3 and 4.2) when $J<0$ (for $J>0$). For convex
boundaries, also $\mu_\text{p/h}/U \simeq \frac{1}{2}\{2g-1-\eta \pm [1-2(2g+1)\eta
+\eta^2]^{1/2}\}$ obtained within meanfield approximation
(cf.\ Refs.~\cite{Mott}) can be expected to provide a reasonable description.

The phase diagram plotted in Fig.~\ref{fig:app}(b) shows: the smaller $n|J|/U$ 
gets, the smaller get the intervals of $\mu/U$ with non-integer filling (i.e.\
outside the MI lobes). In order to reach the strong coupling limit $U\gg n|J|$
at non-integer filling $g<n<g+1$ [where the spin-1/2 description (\ref{eq:Hxy})
with non-trivial polarization applies], the variation of $\mu_i$ 
(i.e.\ of $V_i$) must be smaller than
$\mu_\text{h}^{(g+1)}-\mu_\text{p}^{(g)}\sim2(g+1)w|J|$ over an appreciable
number of sites.
Such a situation, where also the gapped SL phases are supposed to appear, can be
achieved in the center of a shallow trap. Note that the presence of a (shallow)
controllable trapping potential is definitely desirable: tuning its
depth allows to manipulate the chemical potential/filling in the trap center.
With respect to the chemical potential, the gapped SL phases, expected for large
interaction and $\alpha$ near 0.5 or 1.3 [cf.~Fig.~\ref{fig:diagram}], would
appear as incompressible regions at half-odd-integer filling. In
Fig.~\ref{fig:app}(b) we have sketched these phases (shaded in grey); they
show up as ``bubbles'' on the frustrated side between the MI regions.

\section{Experimental signatures {of frustration}}
Experimental signatures of the N\'eel SF are sharp quasimomentum
peaks at $\bp=\bq$ + reciprocal lattice vectors 
{[cf.\ Fig.~\ref{fig:lattice}(b)]}. In the case of spiral order, the
ordering vector $\bq$ can take two different values; when measuring (or already
before) the system will spontaneously choose one of them. For a whole stack of
2D systems, the measurement will average over both quasimomentum distributions,
unless there remains a finite coupling between the 2D-layers establishing the
same order everywhere. Also the predicted downshift of the anisotropy ratio
$\alpha_0$ (where spiral continuously transforms into rhombic-staggered N\'eel
order) with increasing interaction/lattice depth [cf.\ Figs.\ \ref{fig:diagram}
and \ref{fig:app}(a)] can be investigated experimentally.
{The growth of the staggered $\alpha$-domain with increasing quantum
fluctuations is an example for ``order by disorder'' \cite{Reviews}.}
Another measurable consequence of frustration is the extension of
the MI phases to larger values of $|J|/U$ [cf.\ Fig.~\ref{fig:app}(b)]. 
{Due to the lack of long-range order, the MI does not show sharp
peaks in the single- but rather in the two-particle momentum distribution
(noise correlations) \cite{NoiseCorrelation}. This is also true for the gapped
SL phases, being the most striking implication of frustration expected. Thus, in
order to distinguishing the SL from the MI experimentally, one should search for
structures in the momentum distribution beyond sharp peaks. The SL can feature
a pattern in the momentum distribution on the scale of a Brillouin zone (i.e.\
the inverse lattice spacing $\pi/d$), reflecting delocalization of particles on
pairs of neighboring sites forming ``singlets''. Apart from that, single-site
resolved measurements \cite{SingleSite} clearly distinguish between MI and SL by
number fluctuations.}

\section{Conclusion and Outlook}
We have proposed to realize the positive-hopping Bose-Hubbard model with a
system of ultracold spinless atoms in a deep triangular optical lattice dressed
by a rapid elliptical acceleration. Our scheme allows to experimentally
investigate the physics of a frustrated quantum system under the clean and
controlled conditions provided by ultracold atoms. Since frustration is induced
to motional bosonic degrees of freedom, it is experimentally possible to reach
temperatures that are low compared to the energy scales governing the system. 
The model smoothly approaches a quantum spin-1/2 $XY$-model in the deep-lattice
limit of strong interaction.  
In order to draw the phase diagrams shown in Figs.~\ref{fig:diagram} and
\ref{fig:app}(b) we have combined results from different
approaches: (i) numerical simulations applying to the spin-1/2 limit of strong
interaction at half-odd-integer filling (published recently in
Ref.~\cite{SchmiedEtAl08}), (ii) an above-Bogoliubov theory valid in the 
limit of weak interaction, and (iii) analytical strong-coupling as well as
meanfield results for the limit of strong interaction at integer filling.
Expected are superfluid phases showing staggered or spiral N\'eel order,
Mott insulator phases having integer filling, and gapped spin-liquid phases at
half-odd-integer filling.  
We have also described how the positive-hopping regime can be reached
adiabatically, if initially the system is prepared in the usual negative-hopping
ground state. Finally, experimental signatures of the different phases 
have been discussed. In conclusion, using an existing setup the experiment
proposed here can provide novel information about a frustrated quantum system.

We have restricted our analysis to the triangular lattice geometry that we have
implemented already in the laboratory. However, the route described here, namely
(i) realizing a positive-hopping Bose-Hubbard model with ultracold atoms by
dressing the system by a fast elliptical lattice acceleration and (ii)
approaching the physics of a spin-1/2 $XY$-model in the limit of strong
interaction, applies equally to other two-dimensional non-bipartite lattices
such as the Kagom\'e lattice. This opens perspectives for interesting future
research.

\acknowledgments
 
We thank R.\ Schmied, T.\ Roscilde, V.\ Murg, D.\ Porras, and I.\ Cirac for
discussions on the spin model. Support by
the spanish MCI [FIS2008-00784, FIS2007-29996-E
(ESF-EUROQUAM project FERMIX)], the Alexander von Humboldt foundation,
Caixa Manresa, and through ERC grant QUAGATUA as well as through the EU STREP
NAMEQUAM is gratefully acknowledged.

\end{document}